\documentclass[pre,%
reprint,
superscriptaddress,
nofootinbib,
amsmath,amssymb,
aps,
longbibliography
]{revtex4-1}
\usepackage{graphicx}
\usepackage{dcolumn}

\usepackage{bm}
\usepackage{caption}
\usepackage{float}
\usepackage{lipsum}
\usepackage{amsmath}
\usepackage{mathtools}
\usepackage{amssymb}
\usepackage{dsfont}
\usepackage{color}
\usepackage{tikz}
\usetikzlibrary{automata, positioning, arrows}
\usepackage{nccmath}
\usepackage{caption}
\usepackage{subcaption}
\usepackage[utf8]{inputenc}
\usepackage{babel}
\usepackage{amsmath,amsthm,amssymb}
\usepackage{graphicx}
\usepackage{physics}
\usepackage{listings}
\usepackage{enumitem}
\usepackage{microtype}
\usepackage[raiselinks=true, linktoc=all,hidelinks]{hyperref}
\usepackage[capitalise]{cleveref}
\newcolumntype{L}{>{$}l<{$}}

\allowdisplaybreaks[1]

\newcommand\symbolwithin[2]{%
  {\mathmakebox[\widthof{\ensuremath{{}#2{}}}][c]{{#1}}}}

\begin{document}

\preprint{APS/123-QED}

\title{Landau Free Energy of small clusters  beyond mean field approach}
\begin{abstract}
    The Landau Free Energy determines the landscape of order parameter  fluctuations that occur in a physical system at thermal equilibrium and, in particular, characterizes the critical phenomena. 
    We propose a semi-analytical approach based on the Fluctuating Local Field method, which allows to estimate Landau Free Energy for small clusters with discrete (Ising model) and continuous (Heisenberg model) order parameter.
    
\end{abstract}

\author{S. Semenov}
\email{roporoz@gmail.com}
\affiliation{Russian Quantum Center, Moscow 121205, Russia}
\affiliation{Moscow Institute of Physics and Technology, Dolgoprudny, 141701, Russia}
\author{A.N. Rubtsov}
\affiliation{Russian Quantum Center, Moscow 121205, Russia}
\affiliation{Lomonosov Moscow State University, Moscow 119991, Russia}

\date{\today}

\maketitle


\section{\label{sec:level1} Introduction}

The Landau Free Energy (LFE) is a powerful tool for describing the critical phenomena and fluctuations of the order parameter in many quantum and statistical systems. This concept was first introduced by Lev Landau in 1937 to describe a second-order phase transition [\onlinecite{Landau1936}]. Near the transition point the free energy potential is expanded in Taylor series up to the fourth order:

\begin{equation}
F_L(T, \eta)=C+a(T)\eta^2+b(T)\eta^4+\ldots
\label{Taylor expansion}
\end{equation}
The potentials of a similar type are used to study various phases and material properties. For example, for the ferroelectric systems the Landau-Devonshire theory provides the phenomenological description of the relationship between the polarization and applied electric field, as well as other properties such as the dielectric constant and the strain of the material [\onlinecite{Devonshire1951, Chandra2007, Hoffmann2019, Marton2017, Li2005, Zhong1994}]. This concept is also important in the study of such statistical systems as molecular magnets [\onlinecite{Kunii2022}], liquid crystals 
and nematic polymers, where it is known as Landau-De Gennes [\onlinecite{Rusakov1985, Stiven, Majumdar2010}] theory. In the theory of collective quantum phenomena, the Ginzburg-Landau potential is widely used for the phenomenological description of superconductivity and Bose-Eishten condensation [\onlinecite{Landau1950, Bailin1984, Grigorishin2016, Griffiths1970}].

It is commonly accepted that the 2-4 expression is in many cases indeed a good approximation for the $F_L(\eta)$ dependence. 
This is e.g. supported by the experimental studies of  ferroelectric and magnetic materials, where the coefficients in the expansion (\ref{Taylor expansion}) can be estimated [\onlinecite{Siannas2022, Kumari2020, Lin2006, Sheng2010}]. 
However estimation of temperature dependence of polynomial coefficients is still a hard task. This dependence shows a non-analiticity at the transition point for bulk materials. For small clusters, the sharp transition is suppressed by fluctuations, but the temperature dependence of Landau coefficients is still non-trivial. 

The Landau Free Energy can be obtained from the numerical simulations. For small systems with a discrete order parameter it can be calculated directly by the enumeration of all states, however the computational complexity grows exponentially with the system size. Methods based on the Monte Carlo approach, coarse-graining and molecular-dynamics simulations apply for a more general situation  [\onlinecite{Geneste2009, Togo2008, Invernizzi2017, Troster2007}]. In particular, the Wang-Landau algorithm [\onlinecite{Wang2001, Kalyan2016, Moreno2022, Brown2005, Chan2017}] proposed about 20 years ago allows to calculate the Landau Free Energy for systems with both continuous and discrete order parameter. 

Much less is achieved than it comes to analytical calculations. Actually the only method at hand is the Mean Field Theory (MFT). In this approach the interacting system is replaced by a non-interacting one in certain effective potential.
MFT-like methods provide a basic tool for describing the collective behavior of   correlated systems. However, the MFT description of critical phenomena is very limited, because the essentially important physics of fluctuations is almost neglected at the MFT level. Furthermore, it is quite hard to construct an improvement of the MFT result, because the theory does not contain an explicit small parameter.

The Fluctuating Local Field (FLF) method was introduced several years ago for more quantitatively accurate estimation of the thermodynamics properties of the classical lattices [\onlinecite{Rubtsov2018}]. Later, the FLF method was applied to fermionic systems such as Hubbard chains [\onlinecite{Lyakhova2022}] and 2D clusters [\onlinecite{Rubtsov2020}, \onlinecite{Lyakhova2022V2}], as well as in the disordered Ising model [\onlinecite{kuznetsova2022fluctuating}]. The main idea of this approach is introduction of additional fluctuating degrees of freedom conjugated with the order parameter. This allows to take into account the most important modes of fluctuations in the system, regardless of their magnitude. The remaining part of fluctuations is threaten perturbatively. At the same time, similar to the MFT, the FLF description greatly simplifies the form of the Hamiltonian of the system, which makes it possible to obtain good quantitative results without using complex technical procedures. 

In this paper, we report that the FLF method allows to reconstruct the landscape of the Landau Free Eneregy of small classical lattices with  a discrete (Ising) and continuous (Heisenberg) order parameter. We compare the results obtained with the numerically exact data as well as with the MFT predictions. We observe that the second-order FLF method outperforms the MFT one, both qualitatively and quantitatively. 

\section{\label{sec:level1} Model and method}

We consider 2D periodical Ising and Heisenberg ferromagnetic periodized $N$-site lattices with the nearest neighbour interaction $J > 0$. Energy of the lattice spin configuration $s$ takes the form: 

\begin{equation}
E \left( s \right)=-J\sum_{<ij>} \left(s_i, s_j \right)
\label{Energy}
\end{equation}

For the Ising model $s_i$ is a scalar taking values $\pm 1$, for the Heisenberg case $s_i$ is a 3-component vector with $\|s_i\| = 1$. We define the Landau Free Energy:
\begin{equation}
    F_L (\eta) = -\frac{1}{\beta} \ln{ \int_{\eta(s) = \eta} e^{ -\beta E \left( s \right)}   d^N s} 
\end{equation}
Here $\beta$ is the inverse temperature,  the integration (summation in the Ising case) goes over all  states with the order parameter equal to $\eta$. The order parameter  is defined as $\eta \left( s \right)  =N^{-1}  \sum_i  {s}_i$; it is a 3-component (single-component) vector for the Heisenberg (Ising) lattice.

In the following  we  consider the system (\ref{Energy}) placed in the external field $h$. The energy of such a system reads: 
\begin{equation}
E_h \left( s \right)=-J\sum_{<ij>} \left(s_i, s_j \right) - \sum_i \left(h, s_i\right).
\label{Energy h}
\end{equation}

The partition function in the external field reads: 

\begin{equation}
    Z_h=\int e^{-\beta\left[F_L(\eta)-N\cdot \left( \eta, h \right) \right]} d \eta.
\label{Partition function LFE}
\end{equation}

The partition function does not change under the shift $\eta + \delta \eta$. The condition $\delta Z_h=0$ leads to the Schwinger–Dyson equation:
\begin{equation}
   \Biggl \langle \frac{\partial F_{L}}{\partial \eta} \Biggl \rangle =  Nh.
\label{SD equation}
\end{equation}

\subsection{\label{sec:level2} Mean field approach}

The main idea of the MFT-like methods is to replace the system (\ref{Partition function LFE}) with a no-interacting one placed in the effective field $\tilde{h}$. Energy of the non-interacting ensemble reads 
\begin{equation}
    E_{MFT}= -\sum_i \left(\tilde{h}, s_i \right) \text{, } \tilde{\eta} =  h + 4J\eta
\label{Partition function LFE}
\end{equation}
This equation defines the self-consistency condition between the effective external field $\tilde{{h}}$ and the average order parameter value $\bar{{\eta}}$. For a non-interacting system, the latter can be easily calculated.

For our purposes it is useful to express this connection by the dependency of the external field $h$ on the order parameter $\bar{\eta}$:
\[
\bar{\eta} = \eta (h + 4 J\bar{\eta}) \Rightarrow h = h(\bar{\eta}).
\]

Now the estimation for Landau Free energy can be derived.
In the thermodynamic limit the average in (\ref{SD equation}) can be approximated by it's saddle-point value at $\eta=\bar {\eta}$: 
The equation (\ref{SD equation}) takes the form:   
\begin{equation}
    \frac{\partial  F_L}{\partial  \eta} (\bar{\eta}) =  Nh( \bar{\eta}).  
\end{equation}
The value of the antiderivative can be found by the straightforward integration:
\begin{equation}
F_L (\eta)  = N \int_{|\bar{\eta}| < \eta} h( \bar{\eta})  d\bar{\eta}.
\label{LFE MF}
\end{equation}

In case of the Ising lattice, all calculations can be performed analytically. The relation between the external field and the order parameter takes the form $h = \frac{1}{\beta} \left( \text{atanh}(\bar{\eta}) - 4J\beta\bar{\eta}  \right)$. The Landau Free Energy is equal to:
\begin{equation}
     F_L (\eta) = \frac{N}{2 \beta} \ln{\left(1-\eta^2\right) + \frac{N \eta}{\beta \cdot \text{atanh} (\eta)} - N J\eta^2}.
\label{LFE MF analytical}
\end{equation}

\subsection{\label{sec:level2} Fluctuating local field approximation}
To introduce the Fluctuation Local Field method, we 
write the partition function for the system (\ref{Energy h}) in the form
\begin{equation}
    Z_h=\int d^3 \nu \int e^{-\beta\left[E_h \left(s \right)-\frac{N}{2 \lambda}\left(\frac{\lambda}{N} \sum_i s_i{N}+ h - \nu \right)^2\right]} d^N s,
\label{pf flf}
\end{equation}
where the integration over the auxiliary variable $\nu$ is introduced. Likewise the order parameter, $\nu$ is a three- (single) component vector for the Heisenberg (Ising) system.  As one can observe, the integration over $\nu$ leads, up to a prefactor, to the expression $Z_h=\int e^{-\beta E_h} d^N s$, corresponding to the partition function of (\ref{Energy h}).


That form of the partition function  could be considered as an ensemble of lattices placed in an external field. Interaction at these lattices is given by the expression
\begin{eqnarray}
\tilde{E}_{h} \left( s, \nu \right)= -  \sum_{i} && \left( \nu, s_i\right) + \frac{    N\left( \nu - h \right)^2 }{2 \lambda} +  \nonumber\\
&& + \underbrace{\frac{\lambda}{2} \frac{\left(\sum_{i}s_i\right)^2}{N} -J\sum_{i,j} (s_i, s_j)}_\text{$W$}.
\label{Energy flf}
\end{eqnarray}
Compared to the system (\ref{Energy}), a new term appears here $\propto
 \left(\sum_{i} s_i \right)^2$ that is responsible for the effective long-range interaction. As one can observe, it disfavours  the spin ordering at the lattice. 
The main idea behind the FLF approach is that the parameter $\lambda$ can be at the particular value $\lambda_0$ taken in such a way that this new ``artificial'' interaction compensates, in average, the $(s_i, s_j)$ term coming from the original Hamiltonian.
At the mean-field level, each spin of the system (\ref{Energy flf}) is subjected to the force $\left(\lambda \frac{N-1} {N}-4 J\right) \langle s \rangle
$ from others. This suggests 
to take $\lambda^{-1}_0 =J \frac{N-1}{4 N}$. 

After choosing this value for $\lambda_0$, we consider $W$ term of (\ref{Energy flf}) as a perturbation of the non-interaction system placed in external field. The free energy of such a non-interacting ensemble  equals
\begin{equation}
F_{\lambda_0}^0(\nu)=-\frac{N}{\beta} \ln z_{\nu}+\frac{N\left( \nu -  h \right)^2 }{2\lambda}.
\end{equation}

The single-particle partition function  $z_{\nu}$ is independent of the field direction and equal to $z_{\nu} = 2 \cosh{\left( \nu \beta \right)}$ and  $z_{ \nu} = 2 \frac{\sinh {\left( \nu \beta \right)}}{\left(\nu \beta \right)}$ for the Ising and Heisenberg model, respectively. 

The thermodynamic perturbation theory in powers of $W$ [\onlinecite{Landau1980}] can be used to obtain further corrections. Whereas our choice of $\lambda_0$ leads to the vanished first-order correction, the second-order gives:
\begin{equation}
F_{\lambda_0}^2(\nu)=-\frac{N}{\beta} \ln z_{ \nu}+\frac{N\left( \nu -  h \right)^2 }{2 \lambda} + \underbrace{\frac12 \|g_{\nu}\|^2  \sum_{i \neq j} \tilde{J}_{ij}^2}_\text{$\langle W ^2  \rangle_{E_0}$},
\end{equation}
where $\|g_{\tilde{\eta}}\|^2=\sum_{ij}\left( \frac{N^2}{\beta^2} \frac{\partial^2 \ln{z_{\nu}}}{\partial \nu_i \partial \nu_j}\right)^{2}$; the quantity $\tilde{J}_{ij}$ equals $J - \frac{\lambda}{N}$ for the nearest neighbors and $- \frac{\lambda}{N}$ otherwise. 


\subsection{\label{sec:level2} Free energy flow}

The partition function (\ref{pf flf}) could be expressed using the FLF in the following way:

\begin{equation}
    Z_h=\int e^{-\beta \left[F_{\lambda }(\eta) - N \cdot \left( \eta ,  h  \right)  + \frac{N}{\lambda} \frac{h^2}{2}\right]} d \eta,
\label{Partition function FLF}
\end{equation}
where we introduce the order parameter variable $\eta = \frac{\nu}{\lambda}$.
One can observe that this expression passes into the definition of Landau Free Energy (\ref{Partition function LFE}) at $\lambda^{-1} \to 0$. 

Let us consider a continuous change of $\lambda^{-1}$  preserving $Z_h$ unchanged. This corresponds to  certain flow of the function  $F_{\lambda}(\eta)$. Our goal is to trace this flow between the point $\lambda^{-1} = \lambda^{-1}_0$, where we have the FLF result, and $\lambda^{-1}=0$. Knowledge of $F_{\lambda}(\eta)$ at the point $\lambda^{-1} = 0$ will give us Landau Free energy we are interesting in.

Consider an infinitesimal shift $\lambda^{-1} \to \lambda^{-1} +\delta \lambda^{-1}$ and  require that the partition function remains unchanged, $\delta Z_h=0$. It gives: 
\begin{eqnarray}
    \int && \left(\partial_{\lambda^{-1}} F_\lambda\right)  e^{-\beta \left[F_{\lambda }(\eta) - N \cdot \left( \eta ,  h  \right)  + \frac{N}{\lambda} \frac{h^2}{2}\right]} d \eta = \nonumber\\
     && =-\frac{N h^2}{2} \int e^{-\beta \left[F_{\lambda }(\eta) - N \cdot \left( \eta ,  h  \right)  + \frac{N}{\lambda} \frac{h^2}{2}\right]} d \eta,
\label{xi linear term}
\end{eqnarray}
which is a linear integral equation for $\partial_{\lambda^{-1}} F_\lambda$. To solve it, let us consider Swinger-Dyson equations for the $Z_h$. Consider an infinitesimal shift of the integration variable $\eta\to \eta + \delta \eta$. That step should also preserve the partition function. Collecting the terms linear in $\delta \eta$, we obtain
\begin{eqnarray}
    \int \nabla F_\lambda \cdot && e^{-\beta \left[F_{\lambda }(\eta) - N \cdot \left( \eta ,  h  \right)  + \frac{N}{\lambda} \frac{h^2}{2}\right]} d \eta = \nonumber\\
    && =  N h \int e^{-\beta \left[F_{\lambda }(\eta) - N \cdot \left( \eta ,  h  \right)  + \frac{N}{\lambda} \frac{h^2}{2}\right]} d \eta,
\label{eta linear term}
\end{eqnarray}
where $\nabla  = \frac{\partial }{\partial \eta}$ having three (one) components for the Heisenberg (Ising) model.   

The second order in $\delta \eta$ gives:
\begin{eqnarray}
    \beta \int && \left(\nabla F_\lambda -  Nh \right)^2 e^{-\beta \left[F_{\lambda }(\eta) - N \cdot \left( \eta ,  h  \right)  + \frac{N}{\lambda} \frac{h^2}{2}\right]} d \eta= \nonumber\\
    &&  = \int \Delta F_\lambda \cdot e^{-\beta \left[F_{\lambda }(\eta) - N \cdot \left( \eta ,  h  \right)  + \frac{N}{\lambda} \frac{h^2}{2}\right]} d \eta,
\label{eta quadratic term}
\end{eqnarray}
where $\Delta=(\nabla, \nabla )$ is the Laplace operator.

Expanding parentheses in (\ref{eta quadratic term})  and using expressions (\ref{xi linear term}) and (\ref{eta linear term}), we obtain the equation 
\begin{equation}
     2 \beta N   \left(\partial_{\lambda^{-1}} F_\lambda\right)= \left(\Delta  F_\lambda\right) - \beta \left(\nabla  F_\lambda\right)^2 .
\end{equation}

This formula can be easily reformulated in terms of the density of states $g( \lambda^{-1}, \eta) = e^{-\beta F_\lambda (\eta)}$, where it takes the form of the heat equation:
\begin{equation}
    \partial_{\lambda^{-1}} g = \frac{1}{2 \beta N}\Delta g.
\label{heat equation}
\end{equation}

\subsection{\label{sec:level2} Optimization problem}

With the condition given by FLF approximation, we deal with a Cauchy problem:  
\begin{equation}
\begin{cases}
\partial_{\lambda^{-1}} g(\lambda^{-1}, \eta) = \frac{1}{2 \beta}\Delta g(\lambda^{-1}, \eta) \\
g(\lambda^{-1}_0, \eta) = e^{-\beta F_{\lambda_0}}.
\end{cases}
\label{Backward Cauchy Problem}
\end{equation}
It is important to observe that the FLF result $F_{\xi_{0}}$ describes the final state of the evolution under the heat equation. Modelling such a backward evolution is an ill-posed problem: direct integration of (\ref{Backward Cauchy Problem}) from $\lambda^{-1}=\lambda^{-1}_{0}$ to $\lambda^{-1}=0$ appears to be very numerically unstable. 

\begin{figure*}[t!]
\centering
\begin{subfigure}{.5\textwidth}
  \captionsetup{justification=centering}
  \includegraphics[width=1.\linewidth]{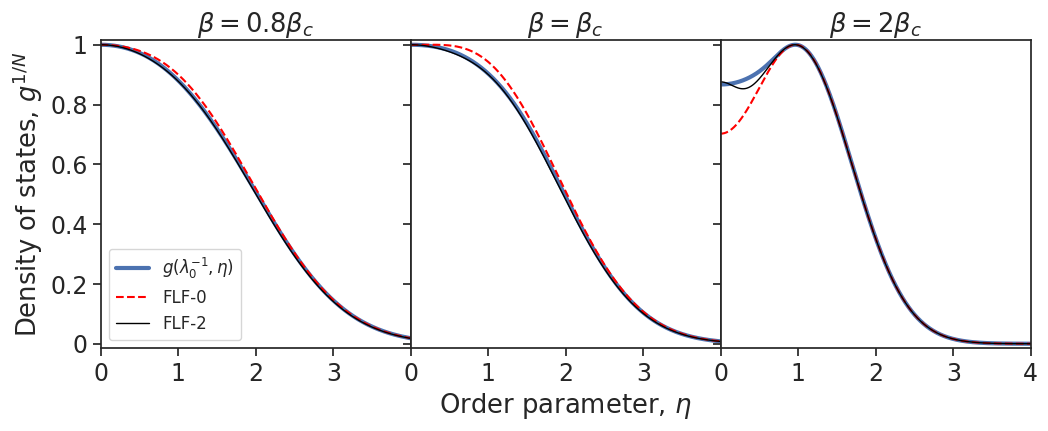}
  \caption{Ising 6x6 periodic lattice}
  \label{fig:sub1}
\end{subfigure}%
\begin{subfigure}{.5\textwidth}
  \captionsetup{justification=centering}
  \includegraphics[width=1.\linewidth]{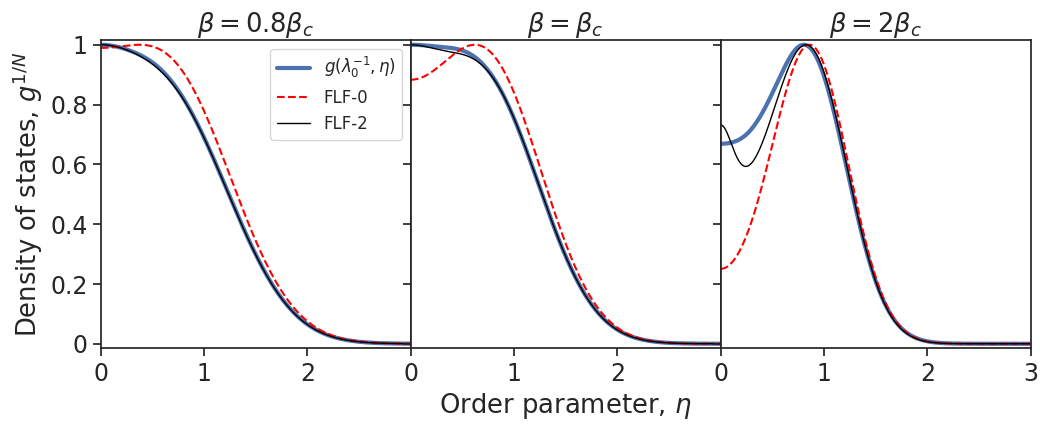}
  \caption{Heisenberg 6x6 periodic lattice}
  \label{fig:sub2}
\end{subfigure}
\caption{Comparison of the evolved reference density of states $g(\xi_0, \eta)$  and the prediction by Local Fluctuating Field method. The curves labeled FLF-0 and FLF-2 refer to diffrent orders of the FLF approximation.}  
\label{fig:dof}
\end{figure*}

To get rid of possible instabilities, we introduce a finite-order polynomial approximation for the Landau Free Energy (\ref{Taylor expansion}): 
\begin{equation}
    F_{a} (\eta) = a_0 + a_2 \eta ^2 + ... + a_p \eta^p
\label{Anzatz}
\end{equation}
Even polynomials of the order from $p=4$ to $8$ were used in practical calculations. For the Heisenberg system, this approximation was used within the range $|\eta| \le 1$; outside this interval the density of states was vanished. For the Ising lattice, $F_a(\eta)$ was defined at the discrete grid $\eta=0, 2 N^{-1}, 4 N^{-1}, ..., 1$ (see the next section for details). 

Given the initial density of states $e^{-\beta F_{a}}$, we solve the heat equation and compare the result $g_a(\lambda^{-1}, \eta)$ with 
the FLF prediction. The coefficients $a_0, a_2, ...$ are adjusted to minimize 
the tolerance $\|g_a(\lambda^{-1}_0, \eta) - e^{-{\beta} F_{\lambda_0} (\eta) }\|$.
A simple gradient-descent method was used to obtain the minimum.



\section{\label{sec:level1} Results}

Here we consider the Ising and Heisenberg small ferromagnetics clusters of the size $4 \times 4$ and $6 \times 6$ with the interaction constant $J = 1$. For these systems, reference data for the Landau Free Energy can be obtained numerically. In the case of 4x4 Ising lattice, we used exact enumeration, in other cases the Wang-Landau algorithm was employed to estimate the density of states.

For the Ising case the order parameter takes discrete values, therefore the initial density of states could be expressed as delta functions sum. The evolved function reads: 

\begin{align*}
 g(0,\eta) &=\sum_{\eta^{\prime}} g\left(\eta^{\prime}\right) \delta\left(\eta-\eta^{\prime}\right) \\
     &\symbolwithin{\Downarrow}{=} \\
 g(\lambda^{-1}_0,\eta) &= A\sqrt{\frac{N}{\lambda_0}}\sum_{\eta^{\prime}} g\left(\eta^{\prime}\right)  \exp \left(-\frac{\left(\eta - \eta ^{\prime}\right) ^2 \cdot \lambda_0 \beta}{2 N}\right)
\end{align*}

For the Heisenberg lattice due to the spherical symmetry of the problem, the three-dimensional heat equation (\ref{Backward Cauchy Problem}) can be reduced to a one-dimensional one by the replacement: $g(\eta) \rightarrow \eta \cdot g(\eta)$.  Since the states exist only for the order parameter $\eta \le 1$,  we suppose $g(0, \eta>1) = 0$. 

In Fig. $\ref{fig:dof}$, we present the evolved density of states $g(\lambda^{-1}_0, \eta)$. The FLF results are compared to the numerically exact reference data.
The $\beta_c$ refers to the critical inverse temperature predicted by the MFT. 
The predictions obtained by the Fluctuation Local Field method are in a good agreement with the functions $g(\lambda^{-1}_0, \eta)$ obtained by evolving the initial condition $g(0, \eta)$.
For relatively high temperatures ($\beta = 0.8 \beta_c$), all curves almost coincide. In this case the FLF approximation perfectly captures fluctuation arising in the system even in the zero order. At low temperatures ($\beta = 2 \beta_c$), the approximation also works well, but a visible difference appears between the orders of approximation. As expected, the second order gives a more accurate prediction, better matching with the density of the states landscape. 

In both cases the FLF method shows a qualitatively correct behavior for the $g(\eta)$ dependence. As $\beta$ increases, the maximum position moves from the disordered state $\eta = 0$ to the polarized one with $\eta = 1$. The only qualitative artifact seen in the FLF-2 curves at low temperature is a local maximum at $\eta=0$. It signals a divergence of the perturbation series at lower temperatures. However, within the range of temperatures considered in our calculation, this issue is not crucial.

\begin{figure*}[!t]
\centering
\begin{subfigure}{.5\textwidth}
  \captionsetup{justification=centering}
  \includegraphics[width=1.\linewidth]{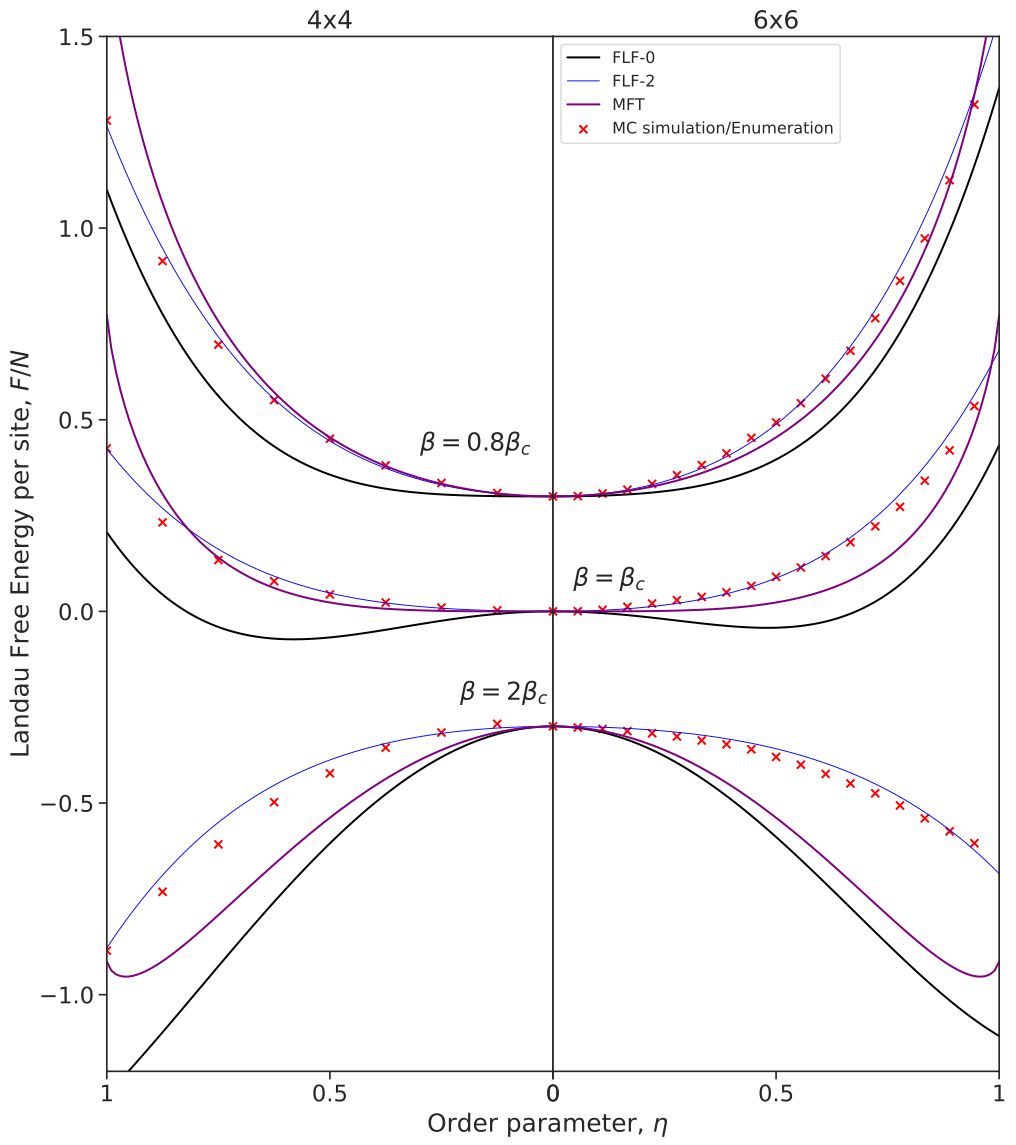}
  \caption{Ising periodic lattices}
  \label{fig:sub1}
\end{subfigure}%
\begin{subfigure}{.5\textwidth}
  \captionsetup{justification=centering}
  \includegraphics[width=1.\linewidth]{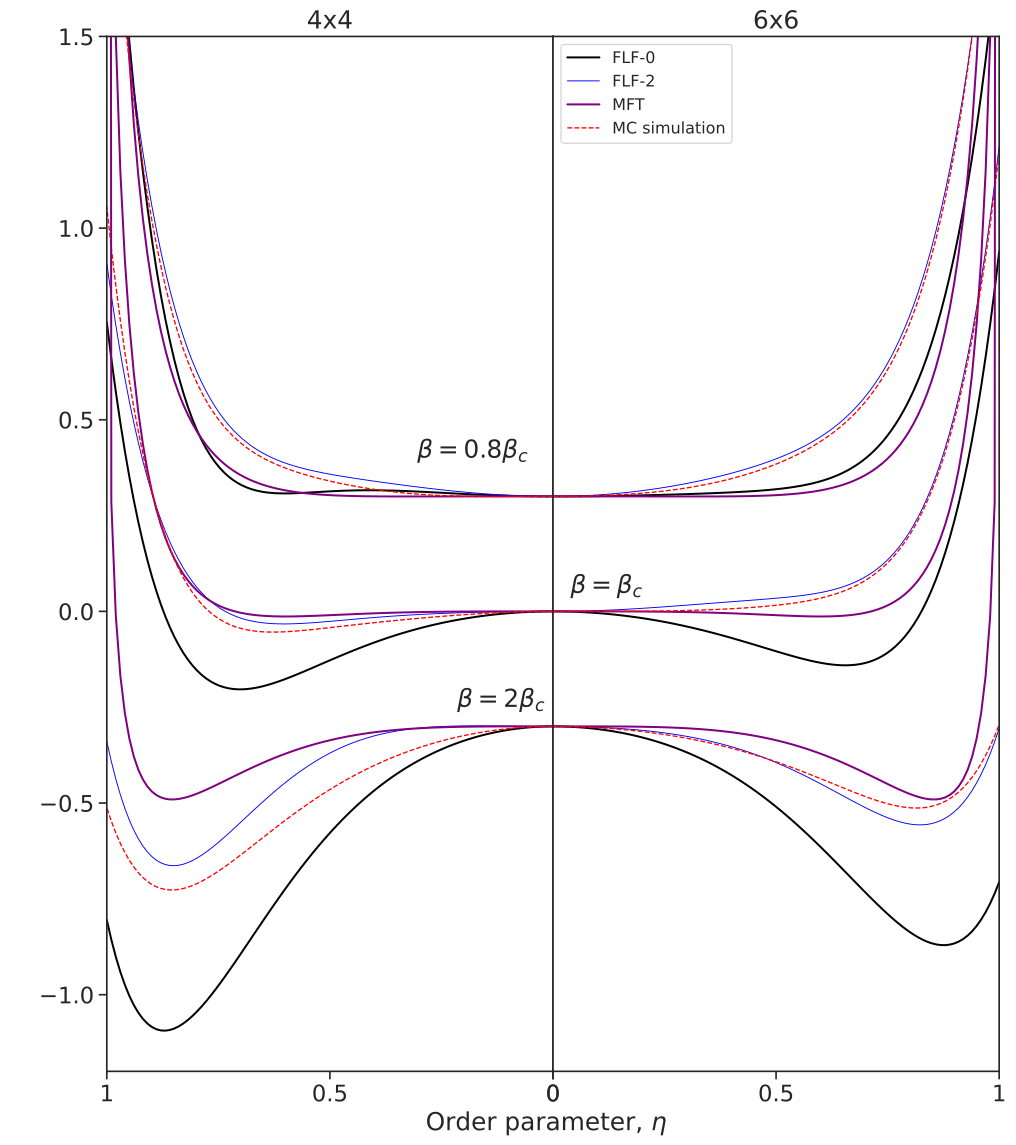}
  \caption{Heisenberg periodic lattices}
  \label{fig:sub2}
\end{subfigure}
\caption{\label{fig:lfe} The Landau Free Energy obtained by the optimization procedure for Ising and Heisenberg 4x4, 6x6 periodic lattices. The resulting curves are compared with the values obtained by numerical simulation. For better visualization, the curves are shifted by a constant for different temperatures.}
\end{figure*}

Fig.$\ref{fig:lfe}$  shows the estimated Landau Free Energy obtained as a solution of the optimization problem. The results are compared with the numerical simulation, as well as with the MFT predictions. 
The MFT curves are calculated using the equations (\ref{LFE MF}, \ref{LFE MF analytical}). As the ansatz (\ref{Anzatz}) in the Ising model we used the $p = 4$  degree polynomial. For  the Heisenberg model the Landau Free Energy curve shows a more complex behavior, so the the $p = 4$ degree polynomial is not sufficient, and we increase its order to $p = 8$. 
The zeroth order FLF and MFT curves show a similar accuracy.  However, there are two quantitative issues about the MFT. First, it gives the results independent of the lattice size.  Second, it predicts two local minima  for the dependence $F(\eta)$ well below the transition point.   However, such a picture is supported by the reference data for the Heisenberg model only. For the Ising case, $F(\eta)$ takes the minimal value at the edge of its domain, $\eta=\pm 1$. It should be also noted that we do not know a simple way to improve the MFT result.
The FLF-0 curve has a dependence on the lattice size, and reproduces the qualitative behaviour of $F(\eta)$ in a correct way. The second order correction greatly improves the FLF results for all types of lattices. An interesting feature of both cases is that for Heisenberg model near the critical point $\beta = \beta_c$, the FLF-0 shows  a minimum at $\eta\neq 0$, but the second order FLF-2 eliminates this inaccuracy. 

\section{\label{sec:level1} Conclusion}

In conclusion, we applied the Fluctuation Local Field method to calculate the Landau Free Energy landscape for classical Heisenberg and Ising small periodical lattices. Our approach establishes an unambiguous relationship between the Landau Free Energy and the equations of the Fluctuating Local Field method. We propose a way to reduce an arising ill-posed backward Cauchy problem for the heat equation to a minimisation problem. This way, we formulate a controllable series of approximations for the Landau Free Energy of small systems.
The scheme is benchmarked by comparison with the results obtained by the numerically exact simulations.

\section*{Acknowledgement}

This work was carried out in the framework of the Russian Quantum Technologies Roadmap.

\bibliography{flf_lfe}

\end{document}